# A Numerical and Efficient Model for Thermo-Field Electron Emission Calculations

Salvador Barranco Cárceles[1,2*], Andreas Kyritsakis[2], Anthony Ayari[1]

[1]Institute Lumière Matière, Université Claude Bernard Lyon 1, Villeurbanne, France

[2]Instute of Technology, Tartu, Estonia

[*]Corresponding author: salvador.barranco-carceles@univ-lyon1.fr, s.barranco.carceles@gmail.com

## Abstract

As electron sources are reduced in size, fabricated in new materials, or pushed for performance, the analytical formulations for thermo-field electron emission, namely Fowler-Nordheim and Murphy-Good equation and their JWKB-based extensions, are increasingly applied out of the range of conditions for which they were derived. We present GETELEC-3, an open-source code that replaces them with a direct numerical solution of the one-dimensional time-independent Schrödinger equation (1D-TISE). Several methods to solve the 1D-TISE have been implemented to test for speed, robustness, and solution's uncertainty. The Noumerov algorithm is found to be quickest, with the solution uncertainty below 1%, and together with an optimised kernel that advances all electron energies in parallel, the cost of calculating the current density is reduced to a few milliseconds – comparable to the JWKB evaluation. To further boost the calculation speed, especially in complicated barriers, a neural network has been trained on the residual of the logarithm of the semiclassical reference and it reproduces the current density better than 0.4% at a fraction of numerical computational cost. The code is developed for metals and semiconductors, accepts arbitrary potential barriers and density of states, and comes with a user-friendly interface and compiled executable to democratise rigorous electron emission calculations and data analysis.

## 1. Introduction

Electron emission from surfaces under high electric fields is at the core of technologies such as high-resolution microscopy and lithography, portable 3D medical imaging, amplifiers for high-speed telecommunications, and as a limiting process in the design and operation of $SF_6$-free power switchers. Its description and analysis have been based in a small set of canonical equations: the Richardson-Laue-Dushman equation for thermionic emission [1], the Fowler-Nordheim for field emission [2], and the Murphy-Good treatment of the Schottky barrier that connects the two through an intermediate regime [3], [4], [5]. Further refinements have been made to improve their accuracy, the regime overlap, and to improve their generalisation for non-planar emitters. However, all of them are based in evaluations of the same integral, in which the transmission probability $D(E)$ is approximated (JWKB) and linearised about one energy [6]. While these approximations have provided invaluable insight into the physics of electron emission for almost a century, they have been also known to have limited application regime and generalisability [7], [8], [9], and yet they are still used in every-day calculations and data analysis.

In Jensen's words, "as demands on electron sources increase, the conditions under which they operate change in a manner that causes simple models and early formulations of the canonical equations to require updating and extension"[10]. Three changes make his point concrete: 1) emitters are now routinely at nanoscale, with apex radii of a few nanometres, where the potential departs from the planar image barrier within the tunnelling distance, some even argue that the one-dimensional time-independent Schrödinger equation (1D-TISE) is not even valid at this scales [11]; 2) the availability of emitting materials have widened: wide-band semiconductors, 2D materials, nanotubes and -wires, or metals with density of states (DoS) far from the free-electron gas approximation [12]; 3) thermo-field electron emission data analysis remains mostly quantitatively with uncertainties that do not allow to differentiate the work function of a plane from that of the next [13]. Even for the planar Schottky barrier, and once an "exact" solution is available, integrating the JWKB-Kemble [14] transmission numerically over energy gives a current density ~6% below the exact one at 300 K over the whole emission range (Fig. 1b). The Murphy-Good equation is 4.7% low at 1.5 V/nm and 7.4% high at 10 V/nm: an error that changes sign with field. At 1000 – 1500 K the error diverges unboundly as $kT$ increases. Therefore, we take on the task of proposing an alternative to analytical solutions that overcomes their current limitations at a competitive cost and helps to the study and design of future electron sources.

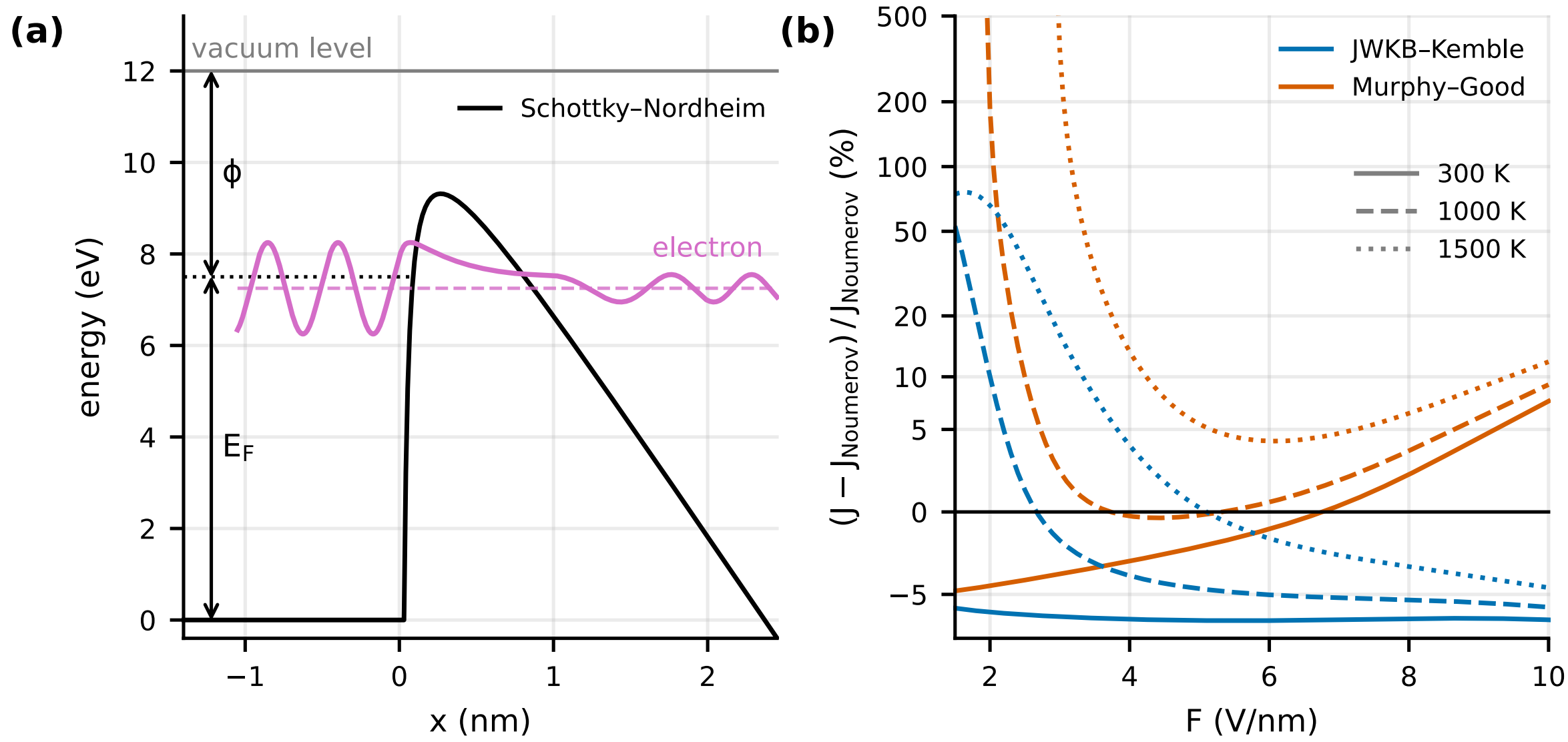

Figure 1 : a) Schematic of the scattering problem: Fermi level $E_F$, work function $\phi$, and potential barrier with apparent height reduction due to image charge. The amplitude of the wave function of the electron (pink) is reduced exponentially during tunnelling and the ratio between the two amplitudes (after/before barrier) gives the transmission coefficient $D$. b) Relative error induced by two analytical solutions to the scattering problem, against the numerical solution, at different temperatures for the Schottky-Nordheim barrier.

Here we present GETELEC-3, a fully numerical method that is robust, versatile, computationally inexpensive, and like its predecessors user friendly [15], [16]. The transmission is obtained by Noumerov (sometimes spelled Numerov) integration of the 1D-TISE though a barrier of any shape, and the current density ($j$), Nottingham heat ($P_N$), total energy distribution (TED), and normal energy distribution (NED) follow from the exact double integral over total and normal energy, for metals and for semiconductors, in any emission regime. The method rests on five contributions that are described in Section 2: (i) a study of several algorithms to solve the 1D-TISE to extract the uncertainty of the transmission solutions ($\delta D \sim 1\%$) and to establish the most efficient of them; (ii) a set of computational measures that make the exact solution cost a few milliseconds per current density while keeping its error below 0.02%, a small fraction of the transmission uncertainty; (iii) inclusion of arbitrary potentials and DoS that can be imported from *ab-initio* methods; (iv) a neural network solver, trained on the residual against a semiclassical reference, that boots the computational speed to barriers with several parametres; (v) a curated guide and user interface to democratise the use of advanced numerical methods. GETELEC-3 is an open-source package (MIT license) and it can be found here.

# 2. Theory and methods

## 2.1 The Scattering problem

Let's imagine an electron travelling from left to right and let's assume the electron can be described as a plane wave with amplitude $A$ (Fig. 1a) as in the free electron model. As the electron travels to the right, it encounters a barrier such that its kinetic energy is lower than the potential of the barrier. Under the current understanding we have about quantum tunnelling, and the problem is not close to be settled, one part of the wave is reflected with amplitude $B$ and another part is transmitted. The transmitted part, under the barrier, experiences an exponential decay of its amplitude and when it exits it does so with a smaller amplitude $C$. Taking the ratio $C/A$ (calculating the ratio of the quantum currents or calculating the Wronskian) one can compute the transmission coefficient $D$ for an electron with a given energy passing through an arbitrary potential barrier. This process has been previously described at length by many authors and we won't go any further, we recommend for further lecture: Gomer's book (Chapter 1) [17] where a clear geometrical interpretation is presented and Griffiths' book (Chapter 8) [6] where the WKB approximation and Airy solutions are very nicely described.

The transmission coefficient is, thus, obtained by solving the 1D-TISE to know the wavefunction of the electron ($\psi$) across the whole domain and obtain its amplitude. This is a lengthy and costly process, especially if it has to be repeated for multiple energies (often every 10 meV or less, over 1 – 2 eV), and it is the reason why analytical solutions were developed. Several methods have been proposed over time to solve the 1D-TISE, with a wide range of efficiency and characteristics (e.g., the transmission matrix method (TMM) can be applied from left to right and right to left and the solution, within the solution uncertainty, doesn't change – this can be useful for certain scenarios). The key and costly magnitude to find is the quantum current

$$j = \frac{\hbar}{m} \Im(\psi * \psi') \quad (1)$$

where the wave function $\psi$ is defined by the Halmitonian

$$\left[ -\frac{\hbar^2}{2m} \frac{\partial^2 \psi}{\partial z^2} + U(z) \right] \psi(z) = E\psi(z) \quad (2)$$

We have tested 8 methods: JWKB-Kemble (numerical integration), Murphy-Good (analytical), TMM, Ordinary Differential Equation solver (ODE, from Scipy), Initial Value Problem solver (IVP, from Scipy), Runge-Kuta 45, Non-Equilibrium Green Functions (NEGF), and Noumerov [18], [19] – we only provide reference for the Noumerov algorithm because it seems to be an obscure method despite its potency. A description of the Noumerov methods follows.

For each energy $E$ the transmission is obtained from the stationary Schrödinger equation along the surface normal,

$$\psi''(z) = q(z)\psi(z) \quad (3)$$

$$q(z) = \frac{U(z) - E}{\hbar^2/2m} \quad (4)$$

where $U$ is the potential, $m$ is the free electron mass, $\hbar$ is the reduced Planck's constant.

On a uniform grid $z_j = z_0 - jh$ , where $h$ is the step size of the solver (SM Sec. S4), that runs from the vacuum ($j = 0$) into the emitter, the Noumerov method advances the solution with a three-term recurrence

$$\begin{aligned} &(1 - T_{j+1})\psi_{j+1} = \\ &= (2 + 10T_j)\psi_j - (1 - T_{j-1})\psi_{j-1} \end{aligned} \quad (5)$$

$$T_j = \frac{h^2}{12} q(x_j) \quad (6)$$

whose local truncation error is $-h^6\psi^{(6)}/240$, i.e. $O(h^6)$. One recurrence per grid point, with no matrix to assemble or invert, is what makes the method cheap even if the truncation error is at least two orders lower than the other methods.

The integration runs from the vacuum into the emitter. In the vacuum, beyond the barrier, only the transmitted wave exists, so the solution is seeded there as a purely outgoing wave; inside the barrier the physical solution grows exponentially towards the emitter, so it is the dominant solution in the direction of integration and round-off does not contaminate it. The seed is a WKB wave

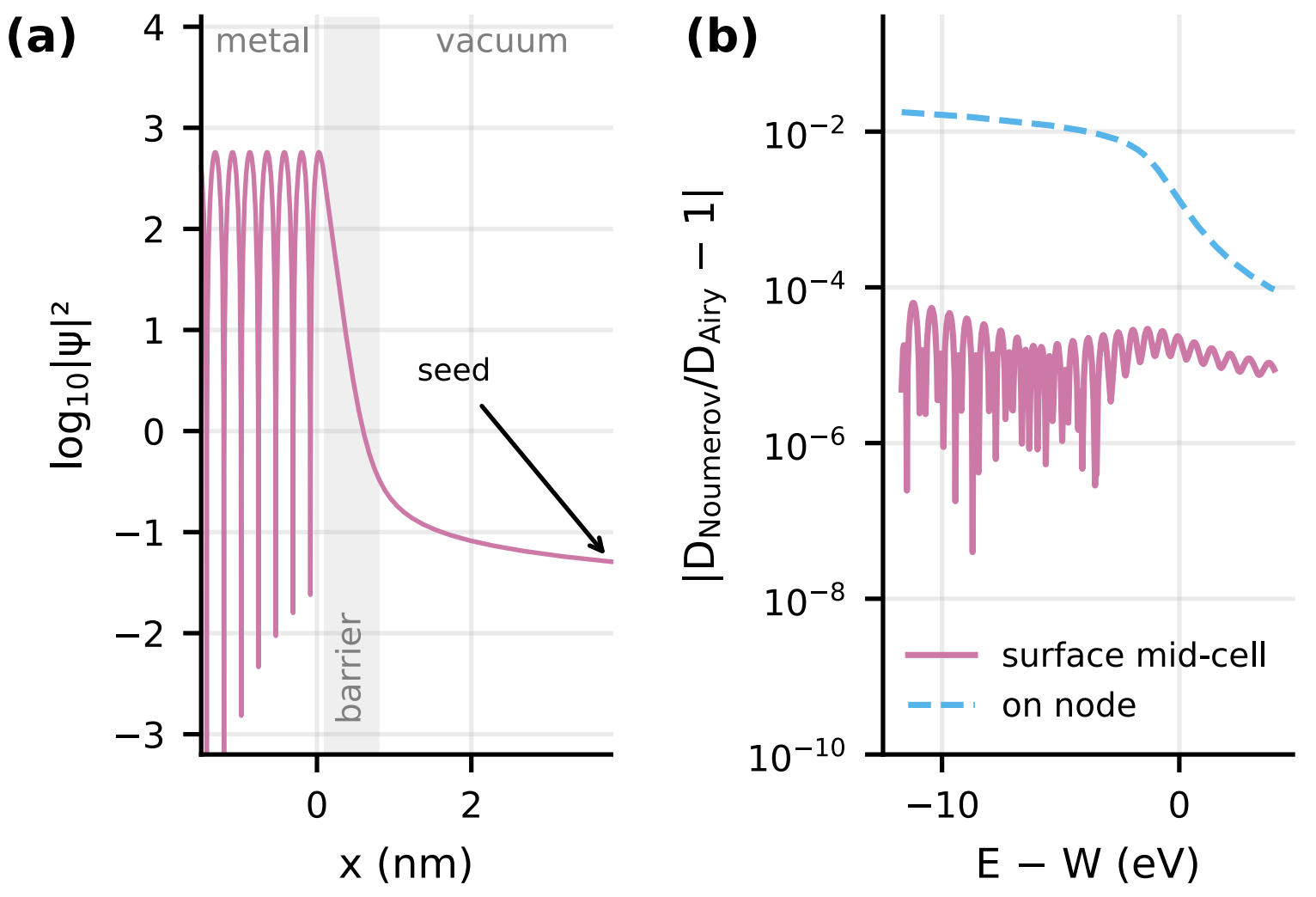

(c)

| method | $D(E_F)10^{-7}$ |
|---|---|
| WKB (analytical) | 1.3023 |
| WKB (numerical) | 1.3023 |
| ODE | 1.1434 |
| IVP | 1.3352 |
| TMM | 1.3222 |
| Noumerov | 1.3224 |
| NEGF | 1.3265 |
| RK45 | 1.3238 |
| Neural Network | 1.3223 |

Figure 2 : a) $log_{10}|\psi|^2$ of the scattered electron at $E = E_F$, for the triangular barrier (forbidden region in grey) and with $\psi$ normalised to the outgoing flux. The integration of the Noumerov method starts from the vacuum side, where the seed is, to the left. b) Relative deviation of the Noumerov solution against the exact Airy solution for the triangular barrier, with the surface placed between two nodes (pink – solver default) and on a node (dashed blue). c) Transmission coefficient at $E_F$ calculated with various algorithms: $E_F = 7.5\ eV, \phi = 4.5\ eV, F = 3$ V/nm, for the Schottky-Nordheim barrier.

$$\psi(z) = k(z)^{-1/2}\exp\left(i\int^{z} k(z')dz'\right) \quad (7)$$

$$k(z) = \sqrt{(E - U(z)/(\hbar^2/2m)} \quad (8)$$

imposed at the first two grid points, with the phase over the first step integrated by Simpson's rule. It carries unit probability current and, unlike a plane wave seed (which can be used as well), remains accurate where the potential is still falling at the end of the domain. Inside the emitter, where $U = 0$, the solution at the last two nodes $z_a$,$z_b$is projected on the incident and reflected waves, $\psi = Ae^{ik_m z} + Be^{-ik_m z}$ with $k_m = (E/(\hbar^2/2m))^{1/2}$, by the closed-form inverse of the $2 \times 2$ system. The transmission is the ratio of the outgoing to the incident current,

$$D(E) = \frac{1}{k_m|A|^2} \quad (9)$$

We have noticed that the Wronskian is numerically more robust than the subtraction method due to rounding errors because of the limited resolution of the data type. Eq. 9 is evaluated as $lnD = -lnk_m - 2\ln\ |A|\text{^}2$: deep in the tunnelling regime $|A^2|$ overflows long before $D$ ceases to be representative, and the logarithmic form returns the transmission down to $lnD \approx -1400$, where the probability itself would underflow at $10^{-308}$.

The same recurrence, keeping every grid point, provides the wavefunction and the quantum current across the domain (Fig. 2a). As a sanity check we have compared the Noumerov solution against the exact Airy solution for the triangular barrier: over 24 decades the largest relative deviation is < $1 \cdot 10^{-4}$ at 5 V/nm (Fig. 2b). Calculating the transmission coefficient with all the mentioned algorithms and comparing their maximum discrepancy over a wide range of fields (0.01 – 12 V/nm) and energies (0.01 – 16 eV) we compute the uncertainty for the solution of $D < 4\%$ (Fig. 2c). Both the ODE and IVP in gross disagreement with the other methods and we cannot figure out why because they act as black boxes, but if we remove those values the uncertainty in the solution of $D$ drops to $\sim 0.75\%$. So, we establish a discrepancy <1% as the acceptable error when performing code optimisation.

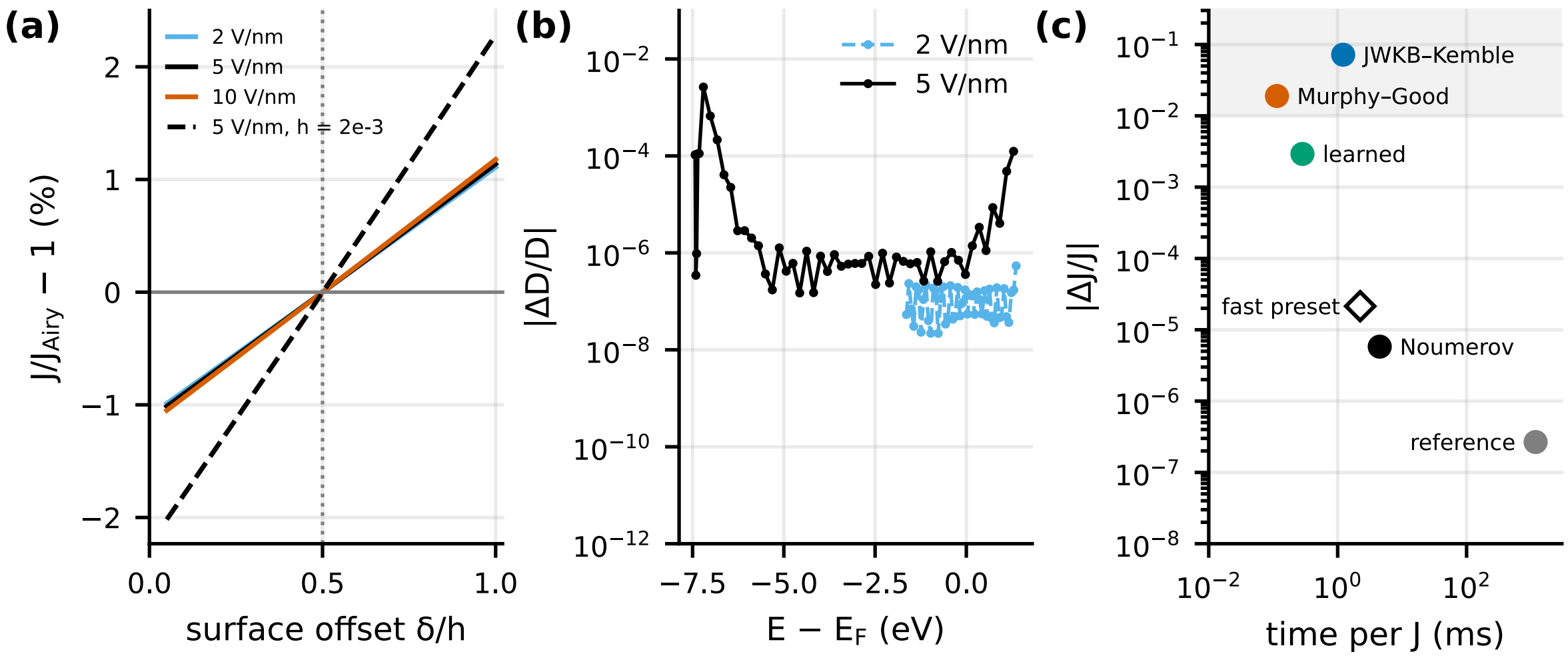

Figure 3 : a) Surface alignment error in j of the triangular barrier against the exact Airy result, as the surface moves across the grid cell ($h = 0.001$nm). The error is exacerbated for coarser grids ($h = 0.002$ nm). b) Error of $D$ when solved on optimised energy grid of 48 + 12 nodes and interpolated in $lnD$. c) Cost and accuracy of one metal current density (5 V/nm, 300 K) for every route in the package, against a converged Noumerov solution ($h = 0.0001$ nm, 20 nm of vacuum), all on the same energy grid – median time. The grey zone shows where the error > 1%.

## 2.2 Algorithm optimisation for speed

The accuracy target follows from the uncertainty of the 1D-TISE solvers (see sub section 2.1), so a numerical shortcut to boost the computation speed is only admissible if the resulting results are within 1% relative error against the reference. Each measure below has been adopted only after its cost had been measured across the thermionic, intermediate, and field regimes. Together, at the default settings, they cost at most 0.3% in $j$ and 0.7% in $D$ – the effect of each optimisation step is illustrated in Figure 3.

**Size of the integration domain.** While during the initial tests a generous integration domain (i.e., -5 – 100 nm) was used to ensure $\psi$ convergence, such a wide domain might be reduced in order to speed up the calculation. Therefore, the outer turning point is located for the lowest energy of each calculation by scanning the potential barrier and starting from the strict planar bound $(W - E)/F$; for sharp tips, because the potential widens, the search is extended from this "flat" point quadrupling the search distance at each step until a point where $E < U$ is found. The domain then extends 3 nm into the vacuum and 0.01 nm (at least two nodes) into the solid. Over 2048 conditions (planar, curved, and triangular barriers with $\phi$ = 2.5 – 6 eV, $F$ = 0.01 – 12 V/nm, $T$ = 300 – 3000 K), this domain reduction changes $j$ by at most 0.0024% and $D$ by 0.025%. This domain shortening is allowed when using a WKB outgoing wave seed, if the plane wave is used the domain after the turning point has to be increased > 20 nm.

**The surface between nodes.** A potential with a discontinuity at the surface (e.g., triangular barrier), that falls at a grid node, induces an error that varies with the position $\delta$ of the surface between its two adjacent nodes, $\propto (2\delta/h - 1)$ (Fig. 3a). With a node at the surface the error is +1.1% with $h = 10^{-3}$ nm; with the surface at mid-cell the error is below 0.0007% against the exact Airy solution for a wide range of fields(Fig. 2b). This process is not done by moving the position of the barrier, that is fixed at $z = 0$, but by displacing the *ruler* that it is created when assembling the potential. If $z_end$ is a whole number of the ruler spacing $h$, the position of the barrier will fall in a node; therefore $z_end$ is displaced by a half to ensure the barrier lands in the mid-point between nodes: $z_{end} = (integer + 0.5) \times h$. For smooth barriers the same shift changes $j$ by $\sim$0.003%.

**A reduced energy set.** $lnD$ is smooth in energy, so it can be solved at a subset of energies and interpolated. The subset is 48 nodes uniformly spaced in the energy grid, plus 12 logarithmically spaced towards the bottom of the band where $D$ falls by many orders of magnitude over a short energy range. A cubic spline of $lnD$ through them is accepted only if it reproduces $lnD$ to <1% at up to 24 probe energies placed mid-way between nodes. Otherwise, every energy is solved. At 5 V/nm this solves 57 of 889 energies and the error is <0.03% where the current flows (Fig. 3b).

**A kernel organised around the data.** Four implementation choices that boost the calculation speed while leaving the results unchanged to round-off error. 1) Only $\psi$ at the last two nodes is needed for Eq. 9, so the wavefunction is advanced in a rolling state and it is never stored: memory falls from $O(N_E N_x)$ to $O(2N_E)$. 2) The Noumerov algorithm is serial for a single energy, i.e., step $j + 1$ needs step $j$ finished first, but across different energies there is no such chain and it can be parallelised. So, by placing the energy grid loop outside and a chunk of energies inside (the number of them is limited by the memory of the computer) where the compiler vectorises it, the kernel can be run 23 times faster on a single core. 3) The coefficients $T_j = \alpha(E - U_j)$, $\alpha = -h^2/[12(\hbar^2/2m)]$, are recomputed rather than read from a $(N_E \times N_x)$ array, which costs more memory bandwidth than the time it saves. 4) The square roots needed for the seed and the projections are evaluated at three points, and not over the entire array. Additionally, to these four implementations, the kernel is compiled with Numba, the rest of the code relies on NumPy and SciPy (highly optimised C libraries), and the kernel runs in parallel over bundles of required simulations.

**Energy grid tailored to emission regime.** The energy grid, instead of expanding uncontrolled, is delimited by two physical criteria: the barrier width and the electron supply for which the relative error is < 1%. When the potential barrier is 3 nm wide or more, tested for the Schottky, triangular, small tip barrier, the barrier can be considered opaque and the emission neglected. The electron supply is cut-off when the supply falls to $5 \times 10^{-24}$ of its value at the Fermi level, which is $\sim 54\ k_b T$ above it. For a metal, with an energy step of 10 meV, the current density stays within 1% wherever $j > 10^{-12} \mathrm{A/cm^2}$. Semiconductors use a 2 meV step, because their grids end at a band edge where the TED has a finite slope.

## 2.3 Arbitrary potentials and DoS

**Potentials.** A potential that has been calculated elsewhere (e.g., a formula, density-functional profile) enters as a function $U(z)$ or a table, and it is solved, integrated, and cached just like the built-in potential barriers. Tabulated potentials are interpolated linearly and the table has to be fine enough (i.e., < 0.01 nm); it has to also extend past the outer turning point, which the code checks for.

**Density of states.** The free electron metal can be replaced by a tabulated density of states $g(E)$, that has been for example calculate with *ad-initio* methods. Each energy is weighted by a number of states it holds to the free electron gas, normalised at the Fermi level,

$$\mathcal{R}(E) = \frac{g(E)/\sqrt{E}}{g(E_F)/\sqrt{E_F}} \qquad (10)$$

$$J(E) = L\mathcal{R}(E) f(E) \int_0^E D \, \mathrm{d}E_z \qquad (11)$$

where $L = qm/2\pi^2\hbar^3$, so that the table acts as the shape correction and $j$ remains comparable to the free electron value. The NED is rebuilt from the same double integral, $NED(E_z) = LD(E_z) \int_{E_z}^{E_{max}} \mathcal{R} f \mathrm{d}E$, accumulated downwards from the top of the grid so that no partial sum loses precession.

## 2.4 Neural Network solver

The case for a learned solver is dimensional: the transmission of the small tip radius barrier, for instance, depends on six variables: $E, E_F, \phi, F, R,$ and $\gamma$; and a table at fixed resolution per axis (as done in GETELEC-2 [16]) multiplies in size with each one (of the order $10^{11}$ nodes for this barrier), whereas the cost of a neural network does not. We train a small neural network per barrier class and keep the Noumerov solver as the ground truth.

The network never predicts $D$. In predicts the residual

$$\mathcal{D}(h, E_F, \phi, F, \dots) = lnD_{Noumerov} - lnD_{ref} \qquad (12)$$

$$lnD_{ref} = -\ln(1 + e^{2G}) \qquad (13)$$

against the JWKB-Kemble transmission of the *same* barrier, with $G$ the Gamow exponential: in closed form, by Gauss-Chebyshev quadrature between the analytic turning points, for the planar potential barrier, and numerically, with the turning points located on a scan of the potential, for a barrier of any shape. The reference carries the exponential dependence, $lnD$ spans by 64 units over the curved barrier (Fig. 4a), and leaves a residual spanning 1.4, with a standard deviation of 0.20. The error budget is explicit since a relative error in $D$ is exactly an absolute error in $ln\mathcal{D}$, so 1% in $D$ means 0.01 in $ln\mathcal{D}$. Learning a correction to a computationally cheap physical baseline rather than the quantity itself is the strategy known in quantum chemistry as Δ-learning [20].

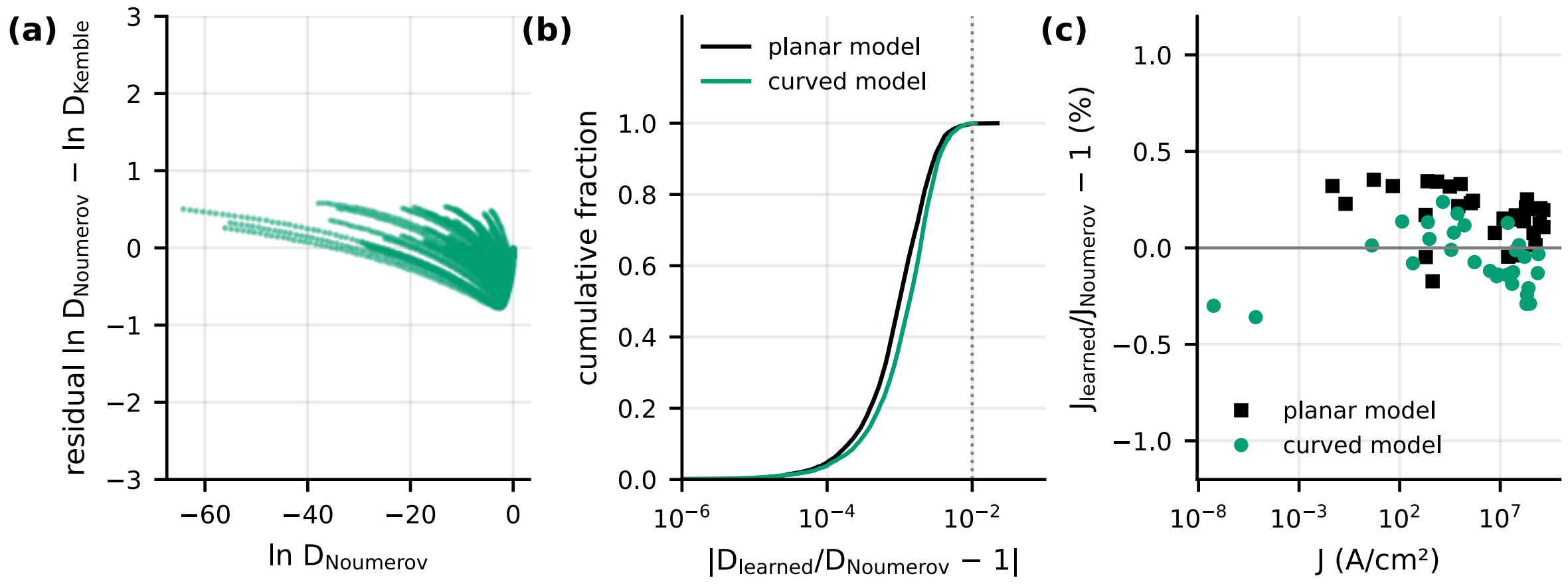

Figure 4: a) Why the residual is learnable: over curved barriers spanning the trained domain, $lnD$ covers tens of e-folds while the residual against the numerical JWKB–Kemble reference of the same barrier stays of order unity. (b) The error of $D$ for both shipped models against an independent, finer Noumerov solution ($h = 0.0001$ nm, 20 nm of vacuum), on barriers not used in training. (c) Error of the current density with the learned solver, over random conditions (the planar model: $E_F, \phi, F, and\ T$; the curved model: also $R$ and $\gamma$).

The networks have three hidden layers of 64 tanh units, such as a unit is the tanh $(f)$ of a physics-derived function $f$, and linear outputs (8 and 15 outputs), and are fitted by full-batch L-BFGS with scikit-learn to Noumerov residuals at 48 random barrier heights between -3 and 7 eV on each of the 2000 planar (96000 samples) and 3000 curved (144000 samples) barriers, drawn over $W = E_F + \phi =$ 9.2–23.5 eV, $F = 0.01 - 12$ V/nm, $R =$ 20–1000 nm, and $\gamma =$ 1–200, with every range extended by 10% so the edges of the training domain are inside the data. The models are stored as plain arrays with the names of their features, reference, and barrier class and need only NumPy to be evaluated.

Both the reference and the forward pass cost time are proportional to the number of energies to be calculated, so the neural network is evaluated on a reduced energy grid (Section 2.2) and $lnD$ splined onto the grid. For a given grid the spline is a fixed linear map, cached as a matrix, so each call is one matrix-vector product, and the forward pass runs in single precision, which changes $\mathcal{D}$ by $10^{-6}$. Outside the training domain the solver hands the calculation to the Noumerov solver.

Against the exact Noumerov solution ($h = 5 \times 10^{-4}$ nm) on 60 barriers per model and 120 heights $W$ each, the median error in $D$ is 0.10% and 014% for the planar and curved barriers respectively. The 99$^{th}$ percentile 0.66% and 0.68%, and the largest 2.2% and 1.1% - in the tails of $D$ where its value is tens of e-folds below its peak (Fig. 4b). These tails error are inconsequential since the current density is within 0.36% in every case over 30 random barrier configurations for each model (Fig. 4c). A current density costs 0.13–0.26 ms, several times less that the default Noumerov solver (Fig. 3c).

### 2.5 Code democratisation and pedagogy

Regardless of how accurate, versatile, or fast a code is, its widespread usage is not so marked by its features, but by the barriers and learning curve that a user encounters when using the code for the first time and most likely in the solitude of their office. Besides, a certain level of expertise in coding or using a particular language cannot be assumed. Therefore, and in order to tackle this issue, we have made GETELEC-3 available in three modes: 1) An executable ready to use app, with an easy-to-use user interface. The degree of freedom that a user can enjoy is limited to the presets available in the GUI; however, calculations of $I-V, I-T, P_N, TED, \text{and } NED$ are available for metals and semiconductors. Likewise, the user can perform data fitting ($I-V$ and $TED$ for metals) just by dropping the .txt/.csv data file in the app window. 2) A more experience user can download and install GETELEC from the GitHub repo and easily install it and have it running in a few minutes – detailed instructions are provided in the GUIDE.md. This version of the code provides full versatility, since the user have access to the full range of features that GETELEC offers. It also comes with a notebook with detailed examples about how to use the code and with data to start playing with parametre fitting and DoS simulations. 3) The lighter package is for those who feel at ease with coding and a simple pip install getelec will get them a light version (no data, no notebooks) of the programme.

A good quality documentation is as useful, as it is rare - even for professional software it is not uncommon to find oneself frustrated when reading the documentation. We are not an exception to such common issue, nor have the time to spend hours writing and refining documentation (we prefer to do science), so we have decided to outsource that task to an AI model (Claude Opus 5, high effort – the prompt can be consulted in the Supplementary Materials). Claude was asked to write a full detailed and pedagogical documentation for the code and the GUI. Likewise, it was given a basic set of functions and lines of code and ask to build a clear and comprehensive intro_to_getelec.ipynb for those who first encounter the software. Our task has focused on reviewing the text to ensure the robustness of the code, correct mistakes, and make it clearer.

## 3. Results and Discussion

The implementation of the above steps has resulted in a fully numerical, versatile, and robust code for the simulation and analysis of thermo-field electron emission from metals and semiconductors. Its numerical solver for $D$ has been optimised in such a way that a current density only cost between 2 to 3 times the computational cost of analytical solution, and yet it retains its accuracy well below the uncertainty level for the solution of the 1D-TISE. Figure 5 illustrates the induced error in $D$ when analytical solutions are employed. Two things can be highlighted: 1) The relative error at the Fermi level, the energy around which analytical solutions have been developed to preserve accuracy, is 13% ($F = 3$ V/nm) and it grows to 30% at $F = 7$ V/nm ($E_F = 9.2$ eV, $\phi = 4.5$ eV, for a Schottky barrier); 2) The numerical solution presents a slight curvature over several eV that the analytical solution does not recover and results in the relative error changing sign and raising to almost 40% error for energies that are readily accessible in laboratories. Such systematic and gross errors, in the basic building block of our science, are ought to be erased and not accepted in light of the calculation speed that modern computation and ML enables.

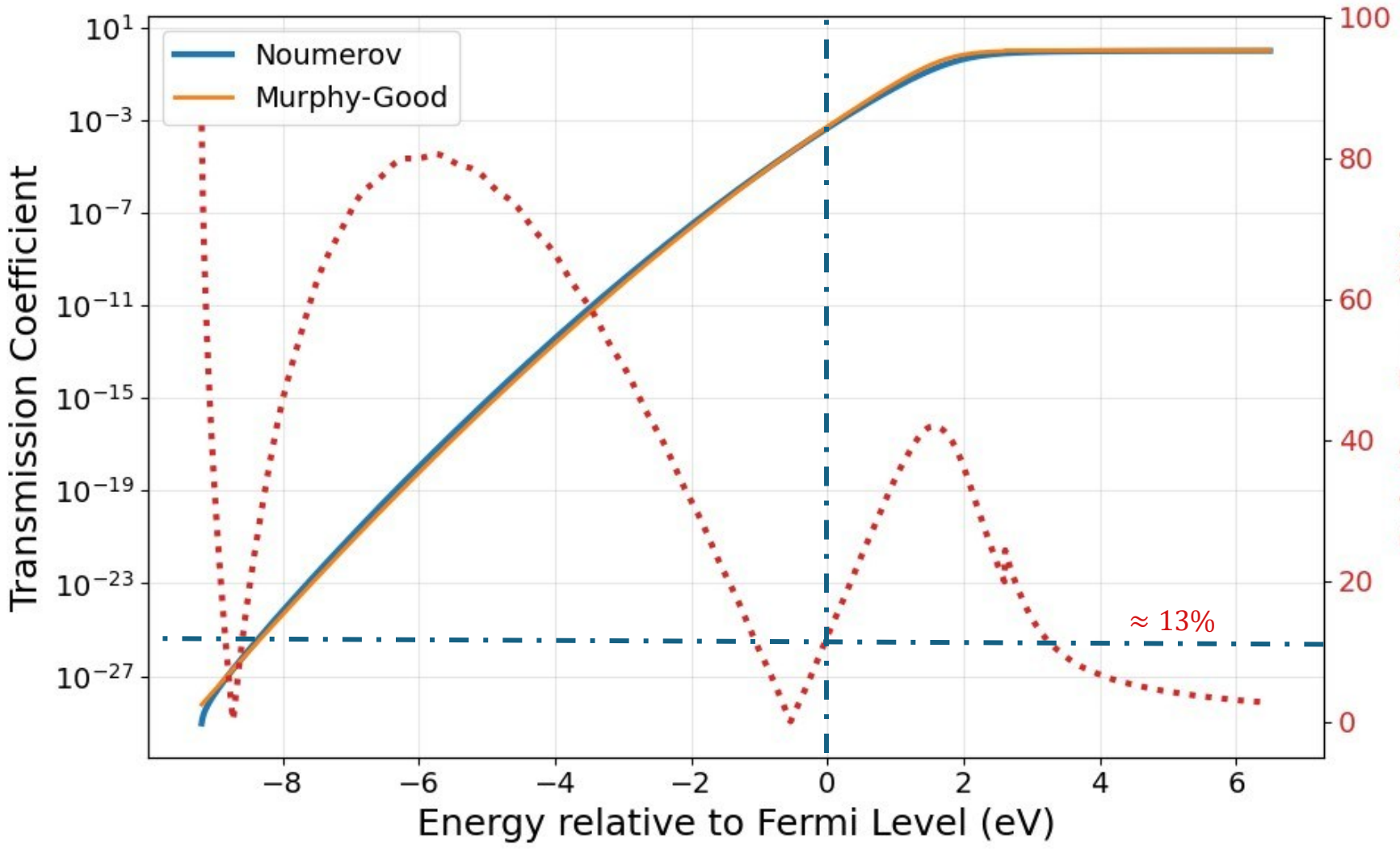

Figure 5: Comparison of the transmission coefficient $D$ calculated analytically (orange) and numerically (blue), with relative error (dashed red). For a field $F$=3 V/nm the error at $E_F$ is approximately 13% and it grows up to 30% for $F$=7 V/nm.

The code is general and applicable to all thermo-field electron emission regimes, without the need of connecting equations or regime change analysis, and accepts imported potential barriers and density of states – certainly it can also be used as a boundary condition for more sophisticated programmes such FEMOCS. For applications where every fraction of a ms counts, the neural network pushes the running time to that of analytical solution but retaining the accuracy of the numerical. We reckon that the same procedure that we have followed in Section 2.4 can be applied directly to current density; thus, saving the user the running of a matrix multiplication and a double integral. Such implementation will lower the running time to parity of that of the Murphy-Good current density – we have not tested that and it is left for a user that is very concern about time management. Certainly, we see an opportunity to push computation performance to explore more complex phenomena (e.g.; plasma initiation and propagation) by the implementation of ML and AI algorithms,

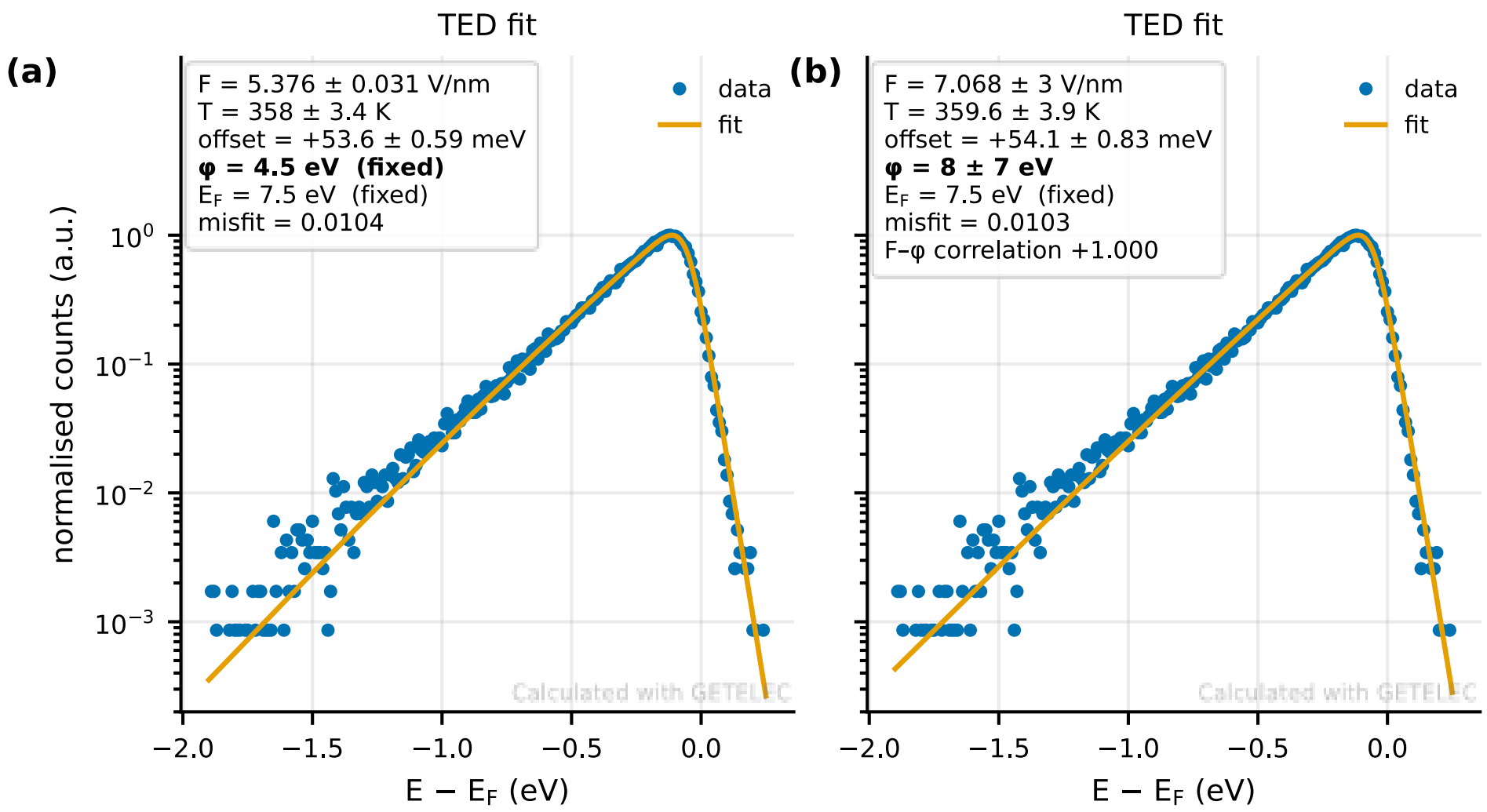

Figure 6:Fit of the same total energy distribution: a) $E_F$ and $\phi$ are fixed at 7.5 and 4.5 eV respectively, leaving the $F$, $T$, and the offset as fitting parametres. b) Like a) but with $\phi$ as a fitting parameter. Visually they both look like good fittings, but when $\phi$ (or $F$) are left as a free fitting parameter they exhaust the fitting range going to its maximum values (e.g., 8 eV here – see b) and their uncertainty explodes. This is due to the fact that the solutions are degenerate and shows that no reliable data can be obtained from TEDs or I-Vs.

as long as they are trained in accurate and trusted data.

GETELEC-3 provides as well the possibility to perform data analysis of $I-V$ and $TED$ without the need of implementing complicated fitting algorithms – just let your data drop at the window and click fit. However, we warn that these analyses are unreliable and in the best of cases their uncertainty does not allow to differentiate between one work function and the next. As a sanity check for unaware users, the code not only fits the data, but also calculates the uncertainty of the extracted fits. Figure 6 illustrates this point: the same data is *visually* well fitted but one gets very different results just by fixing or letting free a parameter (i.e., $\phi$). Regardless, the fitted data can be saved as .txt or as a .png/.svg figure.

## 4. Conclusions

Analytical solutions to the tunnelling problem have helped to develop scientifically and technologically thermo-field electron emission for almost a century. While they will be still useful as a pedagogical tool that any researcher and technologist should master – they provide a clear vision and build intuition of the machinery behind electron tunnelling – their use is no longer justified on the pretext of computational economy. We show that: 1) analytical solutions result in big errors even in the simplest realistic barrier and 2) the computational cost is no longer an excuse. In fact, there is still room for improvement with the ingenious implementation of ML methods.

As the novel electron sources depart from the classical ones (e.g., atomically sharp tips, wide band gap semiconductors, resonant barriers...)

and as we probe more and more the limits of our understanding (studies in the initial velocity of the electron, the effective mass boundary condition…) we need robust and trusted tools we can rely upon, so we do not go astray in the search of new knowledge. We propose GETELEC-3 as one of such tools. Moreover, these tools have to be easy to use by a wide spectrum of expertise, from physical mathematicians to technologists, in order to close the existing gap between experiments and theory – which is a consequence of the multiple and inadequate models that the community uses for simulation and data analysis. We have used LLM models to write a coherent and pedagogical documentation and user-guide in order to democratise GETELEC-3. To go further, a standalone application is available for those who are allergic to coding.

## Code availability

GETELEC is open source under MIT licence: https://github.com/sbcarceles13/GETELEC, installable with pip install getelec. Archived version: 10.5281/zenodo.23093210

## Acknowledgements

This work has been supported by the Agence National de la Recherche grant ANR-22-CE09-002 by the Estonian Research Council grant KOHTO43.